**Nonlinear Dielectric Microwave Losses in MgO Substrates**

M. A. Hein, Dept. of Physics, University of Wuppertal, Germany.

D. E. Oates, MIT Lincoln Laboratory, 244 Wood St., Lexington, MA 02420-9108, U.S.A.

P. J. Hirst and R. G. Humphreys, QinetiQ, Malvern, Worcs. WR14 3PS, U.K.

A. V. Velichko, School of Electronic and Electrical Engineering, Univ. of Birmingham, Edgbaston, Birmingham B15 2TT, UK.

We have investigated the nonlinear surface impedance and two-tone intermodulation distortion of ten epitaxial $YBa_2Cu_3O_{7-\delta}$ films on MgO substrates, using stripline resonators, at frequencies $f$=2.3−11.2GHz and temperatures $T$=1.7K−$T_c$. The power dissipation decreased by up to one order of magnitude as the microwave electric field was increased to 100V/m for $T$<20 K. The reactance showed only a weak increase. The minimum of the losses correlated with a plateau in the intermodulation signal. The same features were observed for a Nb film on MgO but not for $YBa_2Cu_3O_{7-\delta}$ and Nb on $LaAlO_3$ or sapphire. The anomalous response results from dielectric losses in MgO, most probably due to defect dipole relaxation.

PACS 02.60.Cb, 72.15.Lh, 74.20.De, 74.25.Nf, 74.72.Bk, 74.76.Bz, 84.40.Az, 85.25.Am

**Corresponding Author:** Dr. Matthias Hein,

Dept. of Physics, University of Wuppertal, Gauss-Strasse 20, D-42097 Wuppertal, Germany

Tel. +49 (202) 439-2747, Fax +49 (202) 439-2811, Email: mhein@venus.physik.uni-wuppertal.de

Understanding the nonlinearities in superconductive microwave devices is relevant for applications in communication systems, as narrow-band receive or high-power transmit filters can be susceptible to intermodulation distortion or power-dependent absorption [1-3]. The sources of nonlinearities are not well understood at present, and several groups have reported unexpected features like the recovery of superconductivity [4-8] or non-cubic power dependences of the third-order two-tone intermodulation product (IMD) or harmonic generation [9,10].

We have measured the surface impedance and IMD of ten epitaxial $YBa_2Cu_3O_{7-\delta}$ (YBCO) films on MgO and observed a dramatic reduction of the power dissipation at microwave electric fields $E_{rf} \sim 100$V/m below 20K, which correlated with a plateau in the otherwise cubic IMD signal. We attributed this behaviour at first to the YBCO [11]. However, measurements on a Nb film on MgO revealed the same behaviour, while the anomalies were absent in both YBCO and Nb films on $LaAlO_3$ or $CeO_2$-buffered sapphire. These results prove that the dielectric loss tangent of MgO causes the anomaly. Nonlinear dielectric losses, which can affect any microwave device application, have not been previously reported for MgO in the microwave range.

The 350-nm-thick single-sided epitaxial YBCO films were electron-beam coevaporated at 690°C close to the 1:2:3 cation stoichiometry onto 1cm×1cm×0.5mm MgO substrates [12,13]. For comparison, 400-nm-thick Nb films were magnetron sputtered at ambient temperature onto 2-inch-diameter MgO and sapphire wafers [14] and diced into 1cm×1cm pieces. The films, patterned into 150-μm-wide meander lines, were clamped with two equivalent ground planes to form a stripline resonator [15]. The quality factor $Q$ and resonant frequency were measured at the fundamental resonance $\omega/2\pi=2.27$GHz and four overtones, for input power levels $P_{in}=-85$ to $+30$dBm, or electric fields between the center strip and ground planes of $E_{rf}=0.5-5\times10^5$V/m. $E_{rf}$ can be converted into a total microwave current using $\zeta \equiv E_{rf}/I_{rf} \sim 10^5 \Omega$/m. The effective surface resistance so derived is $R_{eff}=R_s+G\times\tan\delta$, where $R_s$ is the surface resistance of the superconductor, $\tan\delta$ the loss tangent of the dielectric, and $G\sim 0.8\Omega$ a geometry factor [15,16]. Changes of the effective reactance are similarly composed of changes of the penetration depth and dielectric permittivity [17]. IMD measurements with a noise floor of −130dBm were performed with two tones of equal input power, separated by 10kHz. We describe representative results for one YBCO film with $T_c$=90.7K, critical current densities of 1.5 and 5.8MA/cm$^2$ at $T$=77 and 60K, and $R_s$=60μΩ at $T=T_c/2$ and 2.3GHz. Above 20K, $R_{eff}$ remained constant up to $I_{rf}$=0.5A, corresponding to a magnetic flux density of about 15mT. We observed similar results for all other YBCO films on MgO, and for the Nb film on MgO at corresponding reduced temperatures (T<8K).

Figure 1 displays $R_{eff}(T)$ for the two power levels that produce extreme $R_{eff}$-values. Below 20K, $R_{eff}(T)$ decreases at the higher power and approaches 2μΩ. Expecting $R_s \geq 1$μΩ for high-quality epitaxial YBCO films [17], $R_{eff}$(20K) yields an upper-limit for the loss tangent of ≈10$^{-6}$, as expected for MgO at 20K and 2.3GHz [18,19]. However, at low power, $R_{eff}(T)$ increases and becomes dominated by dielectric losses. This is confirmed by the very similar power dependence of $R_{eff}$ for Nb on MgO (inset to Fig. 1). The presence of the anomalies for the two totally different superconductors on MgO, and their absence for the other substrates, reinforces the conclusion that they are caused by dielectric losses in MgO. Using $R_s$~2μΩ for YBCO and $R_{eff}$~20μΩ, we derive tan$\delta$(1.7K)≈2×10$^{-5}$, ~20 times above tan$\delta$(20K). We are not aware of previous reports on tan$\delta$ of MgO below 20K.

The anomalies became weaker at higher frequencies, such that $R_{eff}$ scaled like $\omega^{\kappa(P)}$ with an exponent $\kappa$~1.0 (1.4-1.5) at low power and $T$=1.7K (5K). The minimal $R_{eff}$ values scaled quadratically with frequency, independent of temperature, in accordance with the surface resistance of the cuprates [17]. The weaker frequency dependence at low power confirms once more the increasing power dissipation in the dielectric.

Figure 2 relates $R_{eff}$ of the YBCO sample to $\Delta X_{eff}$ and the IMD signal at 5K and 2.3GHz. $R_{eff}$ passes through a minimum at −20dBm, where it has dropped to about 25% of its low-power value. $X_{eff}$ is constant at low power and increases smoothly above −30dBm. Pair breaking in the superconductor or changes of the permittivity obviously cannot explain the anomaly, while the increase of $R_{eff}$ and $X_{eff}$ at high power is attributed to the nonlinear response of the cuprates [17]. The IMD signal at low power displays approximately the expected cubic power dependence [17,20,21]. It passes through a plateau in the region where $R_{eff}$ decreases, and rises again more steeply where both $X_{eff}$ and $R_{eff}$ increase.

We model the impedance of MgO by defect dipole relaxation [22,23]. The same mechanism was attributed to the linear, $T$-dependent, microwave response of LaAlO$_3$ [24]. In Fröhlich's model, the loss tangent is $[(\varepsilon_r - \varepsilon_\infty)\omega\tau_d]/[\varepsilon_r + \varepsilon_\infty(\omega\tau_d)^2]$, with $\varepsilon_r \sim 9.8$ and $\varepsilon_\infty \sim 3$ the permittivities for MgO at low and optical frequencies [18,25]. The relaxation time $\tau_d$ increases for decreasing temperatures, leading to an increase of tanδ in the range $\omega\tau_d < 1$. Our data imply also that $\tau_d(T)$ is explicitly or implicitly field-dependent, e.g., due to dipolar interaction or resonance absorption [22], or heating. We expect $\tau_d(T,E_{rf})$ to depend only on the magnitude of the field, e.g., $\tau_{d0}/\tau_d = 1 + \eta \times e^2/(1+e^2)$, where $\eta$ sets the size of the anomaly and $e \equiv E/E_0$ is the electric field normalized to the scale $E_0$. Our data reveal $E_0 \sim 100$V/m at $T<5$K, increasing to $\sim 30$kV/m at 20K. We also assume $\omega\tau_d < 1$, which is reasonable for ionic crystals like MgO in the low GHz-range [22,23]. The opposite case would lead to a $\tau_d$-dependence of the real part of the complex permittivity, in conflict with the constant $X_{eff}$. The loss tangent is approximately linearly frequency dependent for $\omega\tau_d < 1$, in accordance with the measured frequency dependences. We conclude that $\tau_d(T,E_{rf})$ is the key to explaining the anomaly at low power, while microwave current-driven pair breaking causes the nonlinearity at high power [20]: $f_n = f_{n0} + (1-f_{n0}) \times j^2/(1+j^2)$, with $f_n$ the normal quasiparticle fraction, $j \equiv I/I_0$ a normalized current, and $I_0 \sim 300$mA at $T<20$K. The surface impedance of the superconductor can be derived from $f_n$ in the framework of a two-fluid model.

The model reproduces the essential features observed in experiment (figure 3): The minimum of $R_{eff}$, the constancy of $X_{eff}$, and the IMD-plateau surrounded by cubic behaviour at low and high power levels. The width and depth of the anomaly shrink for increasing $E_0/I_0$-ratios, when the effect of pair breaking increases relative to that of dielectric relaxation [11]. While the extraction of $\tau_{d0}$ and $\eta$ requires separate knowledge of $R_s$ and $\tan\delta$, we can derive estimates by assuming that $\tan\delta$ dominates at low temperatures and frequencies: $\omega\tau_{d0} \sim [\varepsilon_r(R_{eff}-R_s)]/[(\varepsilon_r-\varepsilon_\infty)G]$ and $\eta \geq R_{eff}(E_{rf}=0)/R_{eff,min}-1$. Typical values are at 2.3GHz: $\tau_{d0} \sim 3.1$fs (1.0fs), and $\eta \geq 9$ (3) at $T=1.7$K (5K). Both parameters decrease with increasing temperature. Modeling the data for Nb yielded similar $\tau_{d0}$-values but about a factor of three smaller $\eta$-values. This discrepancy reflects the larger contribution of $R_s$ to $R_{eff}$ due to the much lower $T_c$.

In conclusion, we have observed a strong microwave power-induced decrease of the effective surface resistance of YBCO and Nb films on MgO below 20K, at nearly constant reactance. The third-order intermodulation product displayed a plateau surrounded by almost cubic regions, revealing a clear relation to $R_{eff}$. This anomalous behaviour is caused by dielectric losses in MgO, as concluded from comparisons with YBCO and Nb films on LaAlO$_3$ and sapphire. A phenomenolocial description of the dielectric properties is based on defect dipole relaxation with a temperature and electric field-dependent relaxation time. The low-field loss tangent increases with decreasing temperature from $\sim 10^{-6}$ at 20K to $\sim 2\times 10^{-5}$ at 1.7K. We are not aware of previous reports of nonlinear losses nor dielectric defect dipole relaxation in MgO, but we note reports on self-induced transparency in MgO [26], which were related to the Fe impurities usually contained in this dielectric. Our observations have implications for all microwave applications of MgO, since a nonlinear loss tangent can enhance power dissipation and intermodulation distortion appreciably. Measurements of the complex permittivity of bare and Fe-doped MgO are in preparation.

We gratefully acknowledge contributions from J. Derov, J. Halbritter, K. Irgmaier, N. Klein, P. Lahl, M. Lancaster, B. Moeckley, S.H. Park, N. Newman, and R. Wördenweber. This work has been funded in part by the EPSRC and MOD (UK), and AFOSR (U.S.A.). Part of this material is based upon work supported by the European Office of Aerospace Research and Development, AFOSR/AFRL, under Contract No. F61775-01-WE033.

**FIGURE CAPTIONS**

Figure 1. Temperature dependent effective surface resistance of the YBCO film at $\omega/2\pi=2.3$GHz for two input power levels (−60dBm: hatched squares, −20dBm: dots), corresponding to $E_{rf}\sim 1$ and $10^4$ V/m. Note the logarithmic scales for illustration. The inset shows the normalized $R_{eff}$ versus input power for the YBCO film (diamonds) and a Nb film on MgO (triangles) at $T=5$K. The absolute values of $R_{eff}$ of both samples at 1.7K were identical.

Figure 2. Measured dependence of the third-order intermodulation product (left ordinate, squares), effective surface resistance (right ordinate, dots) and change of effective surface reactance (right ordinate, triangles) of the YBCO film versus input power per tone at 2.3GHz and 5K.

Figure 3. Calculated field dependence of the IMD signal (left ordinate, squares & solid curve), effective surface resistance (right ordinate, dots & dashed curve) and reactance (right ordinate, triangles & dotted curve), in reduced units, for $\omega\tau_{d0}=10^{-5}$, $\eta=3$, $f_{n0}=0.1$, and $E_0/\zeta I_0=3\times 10^{-3}$.

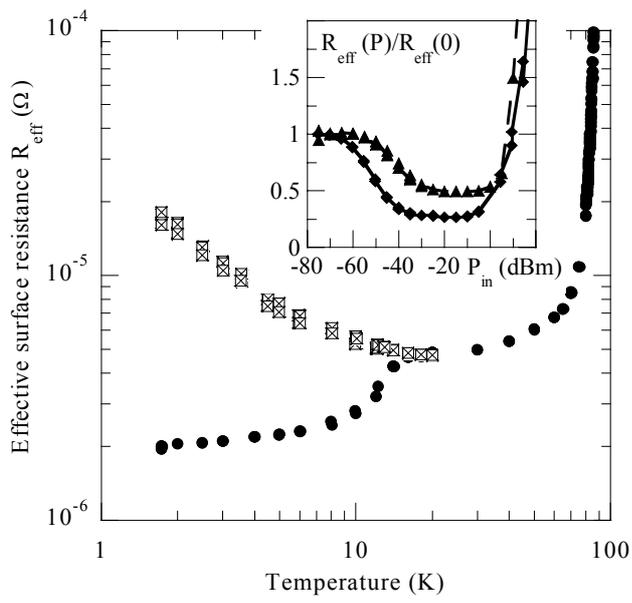

Figure 1, Hein et al., APL

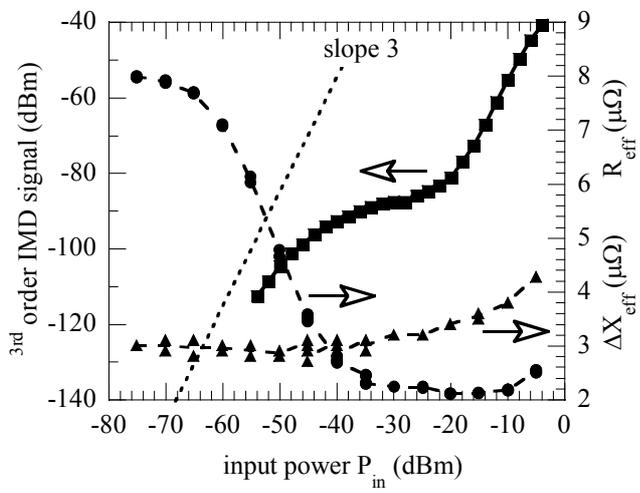

Figure 2, Hein et al, APL

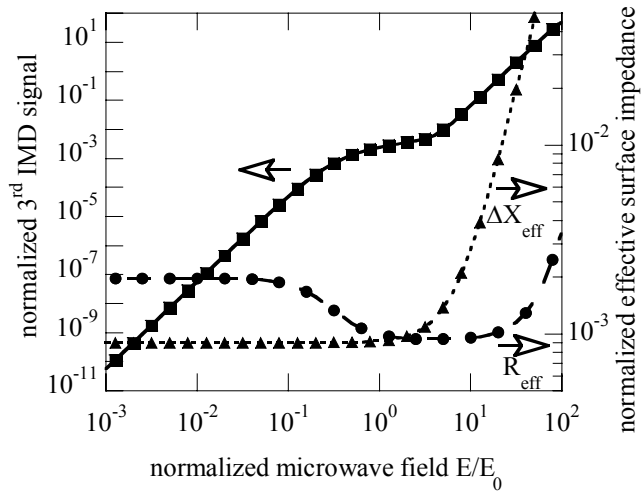

Figure 3, Hein et al., APL